\documentclass[%
 reprint,
 amsmath,amssymb,
 aps,
]{revtex4-1}

\usepackage{graphicx}
\usepackage{dcolumn}
\usepackage{bm}

\begin{document}

\preprint{APS/123-QED}
\title{Nucleation and cap formation on symmetric metal nanocatalysts: A first step towards chirality-controlled single-walled carbon nanotube growth}
\author{Anteneh G. Tefera}
\author{Mogus D. Mochena$^{*}$}%
\affiliation{%
Department of Physics, Florida A and M University, Tallahassee, Florida 32307
}%
\date{\today}
\begin{abstract}
Symmetric minima of surface potential energy of a nanocatalyst act as nucleation sites for chirally selective initial growth of 
single walled carbon tubes at low temperatures. The nucleation sites are sites of maximum coordination number 
of the adsorbed carbon. We show this using the five fold symmetry of a pentagonal pyramid of an icosahedron. 
Initial zigzag structure from nucleation sites results in formation of hexagons and pentagons that result 
in anomalous cap formation. Possible cap lift off mechanism is discussed.

\begin{description}
\item[PACS numbers]
61.48.De, 81.10.Aj, 81.16.Hc, 31.15.A-
\end{description}
\end{abstract}

\pacs{Valid PACS appear here}
\maketitle

The controlled growth of single-walled carbon nanotubes (SWCNTs), i.e. with predetermined diameter and chirality angle, 
remains one of the holy grails of nanotube research despite two decades of substantial progress in their 
synthesis~\cite{1,2,3,4}. The chirality indices (n,m) characterize the diameter and the chirality angle, 
and determine whether a SWCNT is a metal or a semiconductor. At present SWCNTs with a mixture of 
chirality indices are synthesized in a variety of ways~\cite{5}, and growing them with specific chirality as desired is
crucial for their applications in future-generation electronics~\cite{6}. At the core of the problem is the complex
nature of the growth conditions. While it has been possible to observe some aspects of the growth 
in-situ~\cite{3,7,8}, the atomistic dynamics that takes place is beyond the reach of experimental 
observation. Computationally the non-equilibrium processes have been studied with different techniques including 
density functional, tight binding, and classical molecular dynamics calculations~\cite{9,10, 11,12,13,14}. 
These studies have shed light on the growth mechanisms in general, but few have looked into how the chirality and diameter 
are set during the growth. Reich et al showed in the limit of large catalysts how certain pre-assembled carbon caps of root growth are preferred 
during nucleation due to their epitaxial relationship with solid catalyst surface~\cite{15}. Zhu et al, however,  
questioned the rigid epitaxial relation used in the model at the interface between the cap and the catalyst~\cite{16}. 
The work of G$\acute{o}$mez-Gualdr$\acute{o}$n et al 
also shows how the growing nanotube affects the shape of the catalyst~\cite{17}. Nevertheless 
the formation of the cap, and the size of the catalyst are agreed upon generally to be major deciding
factors in chirality selection during the initial stage of growth. The exact way how the cap itself is formed, 
however, has not been shown and has been rarely attempted~\cite{11,18}. In the commonly invoked 
vapor-liquid-solid (VLS) model to show the initial stages of SWCNT growth~\cite{19}, the nanocatalyst 
is first saturated with carbon  atoms and then the carbon atoms form a graphitic sheet on 
the surface and subsequently lift off with 
incorporation of pentagons. During the course of the formation of the 
cap and during the subsequent growth of SWCNT by the incorporation of more carbon atoms at the interface 
of the nanotube and the nanocatalyst, it is argued that ring transformations occur that anneal defects 
into hexagons over the span of the growth time~\cite{20, 21}. This is a 
double aged sword in that while the annealing of the defects to form hexagons is desirable on one hand, 
on the other it shows that at such high temperatures of growth that exceed the Stone-Wales transformations, 
it will be hard to retain the same chirality and local variations are bound to occur. Recently, it has been
suggested because of the chirality instabilities that occur during the growth that it may not be possible 
to achieve chirally controlled growth with nanocatalysts~\cite{22} and that an alternative route 
from nanorings without catalysts may be more realistic. In contrast, Sankaran 
et al claim increasing the percentage of semiconducting SWCNTs to ninety percent in the yield of plasma
enhanced catalytic chemical vapor deposition (PECVD) by tuning the composition of bimetallic catalysts~\cite{4}. 

In this letter we show the possibility of growing SWCNTs with predetermined chirality from
symmetric solid nanocatalyst at low temperature. To our knowledge, the 
symmetry of the nanocatalyst and its effect on chirality selection has not been addressed. We assume
the nucleation sites are symmetric and initate the growth of symmetric SWCNT with pre-determined chirality. 
To test this idea, we chose to work with very small sized nanocatalyst (diameter $<$ 1 nm) because 
such nanocatatlysts lead to diameter-controlled SWCNT growth~\cite{23, 24}. Recently it has been 
shown that carbon atoms grow from nanocatalyst into a nanotube rather than encapsulate it in a fullerene cage due to the 
enhanced energy cost of encapsulation the smaller the nanocatalyst size is~\cite{25}.  

We follow a different simulation approach since we assume multiple adsorptions taking place more 
or less at the same time on the reactive sites of the symmetric faces. Therefore the number of atoms of 
carbon nucleating initially will be at least equal to the reactive sites. This is markedly different from 
the molecular dynamics approach that is often used where one carbon atom or a dimer or a trimer 
is fed to the growth process in periodic intervals or the Monte Carlo scanning and selecting of 
one site at a time based on some criteria. Such sequential simulations of the reactions lead to meandering chains 
of carbon leading to tubular formations of mixed rings of carbon~\cite{9,12,13}.
If the nanocatalyst surfaces are exposed to carbon atoms at the same time, the sequential simulations 
can result in different outcomes than reactions taking place simultaneously.  

The Vienna Ab-initio Molecular Dynamics Package (AMD) is used to perform the calculations with the 
projector augmented wave (PAW) potentials~\cite{26}. The AMD is computationally very intensive, so we chose Fe$_{13}$ as 
our nanocatalyst. Fe$_{13}$ could be dismissed as unrealistic small system, but it has been used to
simulate the growth of SWCNT~\cite{11}, the adhesion strength of monoatomic catalysts of the iron family~\cite{30} and recently to 
study bimetallic catalysts~\cite{31}. From symmetry point of view of our work, it has similar 
properties as the larger more realistic Fe$_{55}$ cluster and thereby one could study the underlying physics
involved in icosahederal nanocatalysts without introducing possible complexities associated with 
larger clusters~\cite{31}.  

The AMD is executed on a picosecond time range while the experimental time 
is on the order of nanoseconds. Therefore, it is criticized as a method that cannot capture the 
experimental processes that yield the final outcome. In this work we argue that the rings and the 
chiralities are set at the initial stages of growth, particularly in the low temperature range of 100 - 500$^\circ$C.
Cantaro et al have grown SWCNTs with thermal chemical vapor deposition (CVD) at temperatures as low 
as 350$^\circ$C~\cite{28}. The possibility of lower temperatures and activation energies than 
thoses of thermal CVD by PECVD has been suggested~\cite{29}.

We performed ground state calculations first~\cite{27}. The resulting optimized structures were then 
heated to finite temperatures by scaling the velocities with temperature. Between the temperature steps, 
a micro-canonical ensemble is simulated. This approach is much faster than the computationally intensive simulation 
of canonical ensemble. 

The icosahedra structure of Fe$_{13}$ is highly symmetric. 
Each atom of  Fe$_{13}$ forms one of the apexes of a pentagonal pyramid when observed from the top of it. 
There are three possible adsorption sites for carbon atoms: the apex, the bridge between the apex and one of 
the atoms at the base of the pentagonal pyramid and the triangular face of the pyramid. The adsorption 
energies were computed for the three possible sites 
at T = 0 K first. The carbon placed on the bridge was pushed to the side and got absorbed on the face and 
resulted in the lowest energy. The triangular face allows the carbon atom to bond with the maximum number 
of Fe atoms (three in this case) and thereby results in lowest energy. The adsorption energy is equal to 7.39 ev 
which is in good agreement with that of~\cite{11}. The relative adsorption energies for the sites with 
respect to this energy are given in Table I.
The question, however, is that are these results for ground state applicable for the finite temperature 
growth of SWCNT? To test their validity, we heated Fe$_{13}$ to 400 - 800 K and found out the icosahedral 
structure of Fe$_{13}$ remained unchanged. Therefore, the carbon atom will see the same surface potential. 
To check whether the kinetic energy of the carbon is large enough to overcome the local minima  
or the diffusion energy barrier on the surface when it moves into the potential well, we heated the optimized 
system, Fe$_{13}$ + C, to 400 - 800 K as well and the results are given in Table I. 
The adsorbed carbon on the face of the optimized structure remained on the face for both cases from T = 0 K, with 
energy difference of 0.14 to 0.52 ev  between the two structures. On the other hand, the carbon attached 
to the apex became slanted at 400 K and 600 K and finally ended over the center of the face at 800 K. 
These results showing that the carbon is still bonded to  the face at low temperatures, despite its 
kinetic energy, validate our assumption that the adsorption sites of T = 0 K still hold for low temperatures.

\begin{table}[b]
\caption{\label{tab:table4}%
Relative adsorption energies (ev). Finite temperature energies are obtained by heating the ground state structures. 
Initial adsorption site is where carbon is placed before optimization.}
\begin{ruledtabular}
\begin{tabular}{ccddd}
Initial Adsorption sites&           
\multicolumn{1}{c}{\textrm{0 K}}&
\multicolumn{1}{c}{\textrm{400 K}}&
\multicolumn{1}{c}{\textrm{600 K}}&
\multicolumn{1}{c}{\textrm{800 K}}\\
\hline
Face&0.07&0.67&1.09&1.57\\
Bridge&0.00 &0.82 & 1.45&2.09 \\
Apex&2.13 & 2.51 & 2.52&1.14  \\
\end{tabular}
\end{ruledtabular}
\end{table}
In our model, carbon atoms flow onto the face from one direction, for instance in PECVD. 
After placing the carbon atoms on the symmetric faces, we determined the next likely adsorption 
site by placing carbon atoms in three different places: i) along the bridge again (the carbon atom bonds with 
four atoms, two from the vertices of the triangular face and two from the adsorbed carbon atoms on adjacent 
triangles), ii) on the apex again, and iii) on top of one of the 
adsorbed carbon atoms on the face. During the adsorption, the carbon atoms tend to redistribute themselves 
or spread out as much as possible to reduce the strain energy. Therefore to promote 
unidirectional growth of nanotubes a geometrical constraint has to be imposed to prevent the spreading out of 
the carbon atoms in all directions on the catalyst surface. Such an assumption has been applied previously by 
passivating the lower portion of the nanocatalyst~\cite{11}. We rule out the spreading out of carbon atoms 
on Fe$_{13}$ icosahedral surface into region below the Fe atoms forming the pentagonal base. The optimized 
structures of the second and third cases violate the geometrical constraint and were ruled out. 
The first case resulted in the lowest energy. The carbon atom 
is pushed outward along with the adjacent carbon atoms adsorped on triangular faces, as seen in the Fig. 1(a). 
This outward push, however, is counterbalanced by adsorption at diametrically opposite site when multiple 
adsorptions occur, and the push outward is not as significant as it looks in the figure. In the process, it forms the 
portion of the zigzag structure that will eventually form around the pentagonal base as seen in Fig. 1(b). 
This configuration is unstable energetically and if left alone 
to relax would optimize into a relaxed structure shown in Fig. 1(c). However, in the presence of carbon atoms 
in the vicinity, the zigzag structure with its kinks is more reactive and forms hexagonal rings as seen in Fig. 1(b). 

\begin{figure}[h!]
 \includegraphics[width=8.2cm, height=12.3cm]{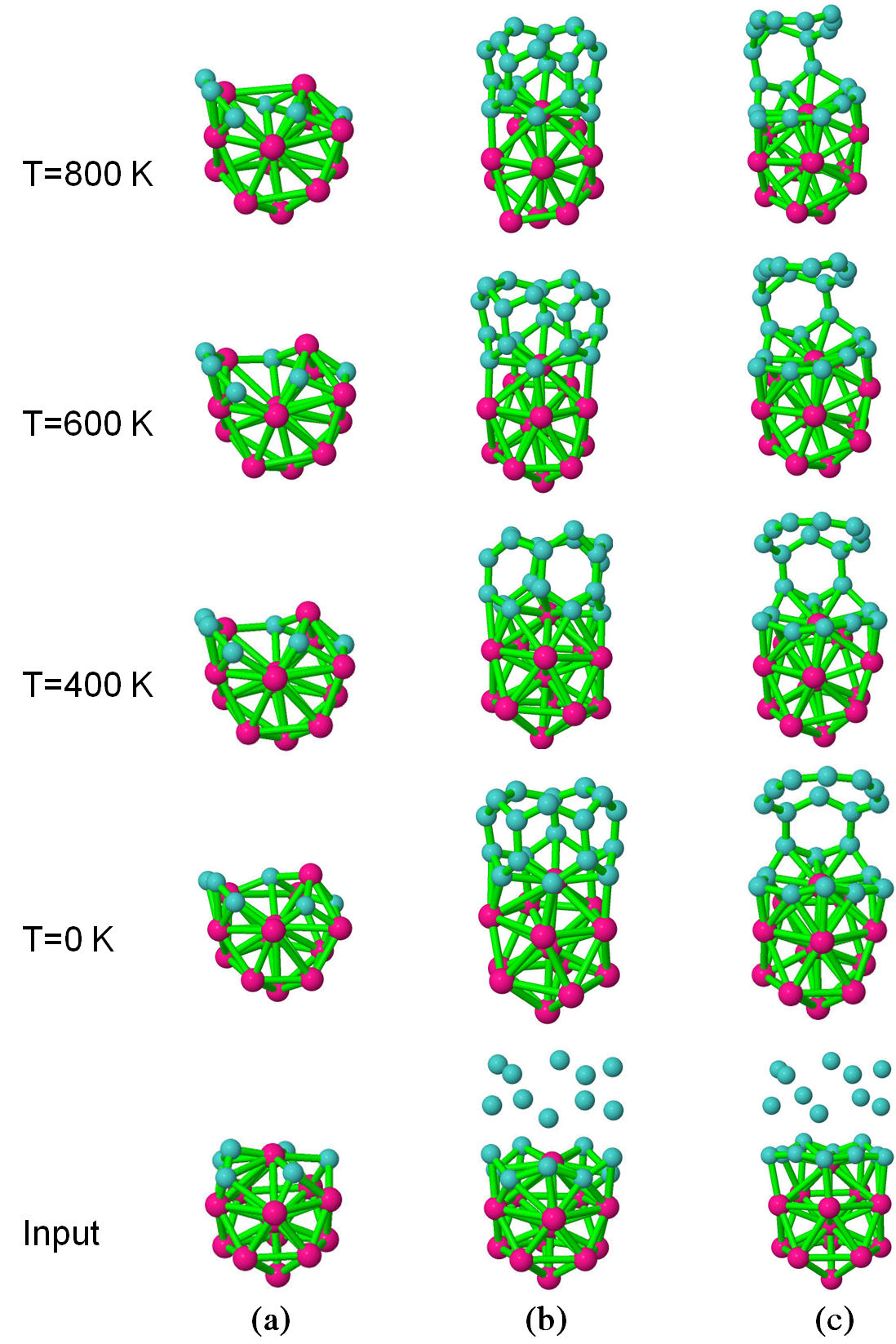}
\caption{(color online) a) Formation of zigzag fragment. b) Initial hexagonal rings formation. 
c) Open structure from relaxed zigzag ring. The T$\neq$ 0 K structures were obtained by heating the T = 0 K ones.}\centering{\label{Fig.1}}
\end{figure}

At T = 0 K we supplied carbon in the form of uniformly distributed carbon atoms, carbon atoms + 
dimers, dimers only, and a ten-atom carbon ring. Three 
of the distributions, carbon atoms, carbon atoms + dimers, and the ring resulted in hexagonal ring formation. 
The ten-atom ring when heated to finite temperature caused the distortion of Fe$_{13}$, therefore, 
we proceeded only with atoms, and atoms + dimers. For brevity, the result for atoms only is shown in Fig. 1(b). 
One could raise the question of supplying randomly 
distributed carbon atoms instead of uniformly distributed ones, but we argue in the tiny diameter range of 
Fe$_{13}$ there is not much volume for randomness and a symmetric supply is a good approximation. 
The relaxed structure of Fe$_{13}$ + 10C was also exposed to the same carbon distribution at T = 0 K, optimized 
and then heated. The optimized structure were open and are shown in Fig. 1(c). The result underscores the fact that 
a zigzag structure with kinks at the onset is critical to the formation of 
hexagonal rings and the icosahedral geometry is conducive to such a formation.
The icosahedral structure of the Fe$_{13}$ remained intact except for the local distortions in the vicinity 
of the adsorption sites at the interface. 
\begin{figure}[h!]
\includegraphics[width=6.5cm, height=8.2cm]{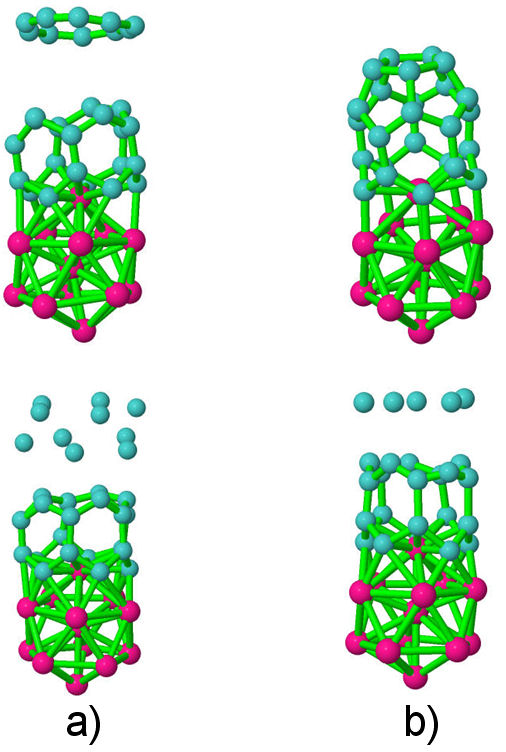}
\caption{(color online) At T = 600 K a) 10-atom ring repulsed from Fe$_{13}$ + C$_{20}$ cluster, b) hemispherical  C$_{20}$ cap formation.}\centering{\label{Fig.1}}
\end{figure}

In Fig. 2, we show the results of simulation of additional carbon atoms. Results for atoms + dimers are 
similar therefore, are not shown for brevity. 
In both cases of the input, a ring as shown was formed and was pushed off instead of forming part 
of the existing structure. There are two triangles formed at the edge of the Fe$_{13}$ + 20C on 
opposite sides as neighboring dangling bonds combined. We broke the bond forming the triangle 
for each triangle and optimized again with the additional ten carbon atoms on the top. The 10-atom 
ring was once again formed and was repulsed. At such low finite temperature, the kinetic energy of 
the carbon atoms is not strong enough to overcome the repulsive barrier to attach to the existing structure. This could 
be explained by the recent findings of charge transfer from the nanocatalyst to the carbon atoms~\cite{32}. 
The carbon atoms adjacent to the iron atoms become negatively charged, and this 
first set of negatively charged carbon atoms  attracts the next batch of neutral carbon atoms by 
polarizing them and integrates them into the structure. This polarization effect pushes the electrons 
to the edge of the nascent tube. These electrons at the edge repulse the $\pi$ electrons of 
the ring of Fig. 2 that was formed during the interaction. The repulsive force dominates over whatever 
remnant attractive force that exists from the dangling bond of the edge carbon atoms trying to bond 
with other carbon atoms on the edge~\cite{33} and the attractive force emanating from the iron atoms, 
resulting in the pushed away carbon ring in Fig. 2. We reduced the number of carbon input atoms to five 
instead of ten to reduce the repulsive force and to see if the attractive force could lead to a closed structure 
instead of an open-ended structure with dangling bonds. The two triangles at the edge were removed by 
breaking the bond between the neighboring carbons at the edge before placing the five carbon atoms~\cite{33}. 
This leads to the hemispherical cap of C$_{20}$  fullerene shown in Fig. 2. The cap is not the usual hemispherical C$_{60}$ 
cap that is often attached to the end of the nanotubes, and does 
not obey the isolated pentagon rule. This anomaly has been recently shown to exist at the end of the (3,3) 
SWCNT~\cite{34}. The ultra-narrow SWCNTs require larger energy of curvature and therefore more 
pentagons that are not isolated are necessary to form a cap. This is counter to the conventional isolated 
pentagon rule, but may be feasible for the ultra-narrow SWCNTs. The larger energy of curvature for these 
ultra-narrow SWCNTs can lead to well defined chirality in contrast to larger diameter SWCNTs whose energy 
of formation are almost identical~\cite{15} leading to a mixture of chiralities. Moreover, in an entirely 
different context it has been shown that SWCNTs grown from nanodiamonds have polyhedron-shaped caps 
rather than the usual hemispherical ones~\cite{35}. Thus the formation of a C$_{60}$ type cap may not 
be the necessary condition for the narrowest SWCNTs. The forces that are available at the nucleation: the 
dangling bonds trying to form bonds with their counterparts and the strong adhesion energy of monometallic 
catalyst such as iron could provide the curvature energy necessary. On the other hand as noted above the 
repulsive force acts as a countervailing force against the attractive force and possibly leads to the cap lift-off. 

In summary we have shown nucleation on reactive sites of symmetric nanocatalyst could lead 
to a well defined chirality of ultra-small SWCNTs at low temperature with icosahedral Fe$_{13}$. The small system 
results in well defined chirality because of the relatively large energy of curvature. A hemispherical cap of C$_{20}$ 
is formed which could lead to the growth of (5,0) nanotube. An electrostatic repulsion counteracting with attractive 
forces of dangling bonds and the nanocatalyst is responsible for cap lift-off. Plasma enhanced chemical vapor 
deposition can supply the carbon atoms or carbon atoms + dimers which can be incorporated through root 
growth mechanism for extended growth.

\begin{acknowledgments}
This work was supported by NSF grant: DMR-0804805. This research was supported in part
 by the NSF through TeraGrid resources provided by NCSA under grant number TG-DMR100055. 

\end{acknowledgments}


\nocite{*}

\begin{thebibliography}{apssamp}  
\bibitem{1}
M.-F. C. Fiawoo, A.-M. Bonnot, H. Amara,1 C. Bichara, J. Thibault-P\'{e}nisson, and A. Loiseau, Phys. Rev. Lett.\textbf{108}, 195503 (2012).
\bibitem{2}
M. Fouquet,  B. C. Bayer, S. Esconjauregui, R. Blume,  J. H. Warner, S. Hofmann 1,  R. Schl$\ddot{o}$gl, C. Thomsen, and J. Robertson, Phys. Rev. B\textbf{85}, 235411 (2012).
\bibitem{3}
R. Rao, D. Liptak, T. Cherukuri1, B.I. Yakobson, and B. Maruyama, Nature Mater. \textbf{11}, 213 (2012).
\bibitem{4}
W.-H. Chiang and R. M. Sankaran, Nature Mater. 8, 882 (2009); D. Dutta, W. -H. Chiang, R. M. Sankaran, V. R. Bhethanabotla, Carbon \textbf{50}, 3766 (2012).
\bibitem{5}
 J. –P. Tessonier and D. S. Su, ChemSusChem \textbf{4}, 824 (2011).
\bibitem{6}
R. H. Baughman, A. A. Zakhidov, and W. A. der Heer, Science \textbf{297}, 787 (2002).
\bibitem{7}
 C. T. Wirth, B. C. Bayer, A. D. Gamalski, S. Esconjauregui, R. S. Weatherup, C. Ducati, C. Baehtz, J. Robertson,and S. Hofmann, Chem. Mater. \textbf{24}, 4633 (2012).
\bibitem{8}
E. Pigos, E. S. Penev, M. A. Ribas, R. Sharma, B. I. Yakobson, and A. R. Harutyunyan, ACS Nano \textbf{5}, 10096 (2011).
\bibitem{9}
 Y. Shibuta, S. Maruyama, Chem. Phys. Lett. \textbf{382}, 381 (2003).
\bibitem{10}
 J. Gavillet, A. Loiseau, C. Journet, F. Willaime, F. Ducastelle, J. -C. Charlier, Phys. Rev. Lett. \textbf{87}, 275504 (2001).
\bibitem{11}
 J. Y. Raty, F. Gygi, G. Galli, Phys. Rev. Lett. \textbf{95}, 096103 (2005).
\bibitem{12}
 Y. Ohta, S. Irle, Y. Okamoto, K. Morokuma, ACS Nano \textbf{2}, 1437 (2008).
\bibitem{13}
 H. Amara, C. Bichara, F. Ducastelle, Phys. Rev. Lett. \textbf{100}, 056105 (2008).
\bibitem{14}
 F. Ding, P. Larsson, J. A. Larsson, R. Ahuja, H. Duan, A. Rosén, K. Bolton, Nano Lett. \textbf{8}, 463(2008).
\bibitem{15}
S. Stephan, L. Li, J. Robertson, Chem. Phys. Lett. \textbf{421}, 469 (2006).
\bibitem{16}
W. Zhu and  A. B$\ddot{o}$rjesson and  K. Bolton, Carbon \textbf{48}, 470 (2010).
\bibitem{17}
D. A. G\'{o}mez-Gualdr\'{o}n and J. Zhao and P. B. Balbuena, J. Chem. Phys. \textbf{134}, 014705 (2011).
\bibitem{18}
Y. Ohta, Y. Okamoto, A. J. Page, S. Irle, and K. Morokuma, ACS Nano \textbf{3}, 3413 (2009).
\bibitem{19}
 R. T. K. Baker, P. S. Harris, F. S. Feates, R. J. Waite, J. Catal. \textbf{30}, 86 (1973).
\bibitem{20}
 E. C. Neyts, A. C. T. van Duin, and A. Bogaerts, J. Am. Chem. Soc. \textbf{133}, 17225 (2011).
\bibitem{21}
 M. Diarra,  H. Amara, C. Bichara, and F. Ducastelle, Phys. Rev. B\textbf{85}, 245446 (2012).
\bibitem{22}
 H.-B. Li, A. J. Page, S. Irle,,and K. Morokuma, J. Am. Chem. Soc. \textbf{134}, 15887 (2012).
\bibitem{23}
 G. H. Jeong, S. Suzuki, Y. Kobayashi, A. Yamazaki, H.Yoshimura, and Y. Homma, Appl. Phys. Lett. \textbf{90}, 043108 (2007).
\bibitem{24}
 N. Li, X. Wang, F.  Ren, G. L. Haller, and L. D. Pfefferle, J. Phys. Chem. C\textbf{113}, 10070 (2009).
\bibitem{25}
 J. C. Burgos, H. Reyna, B. I. Yakobson, and P. B. Balbuena, J. Phys. Chem. C\textbf{114}, 6952 (2010)
\bibitem{26}
 G. Kresse and J. Hafner, J. Phys.: Condens. Matt. \textbf{6}, 8245 (1994).
\bibitem{27}
A. G. Tefera and M. D. Mochena,J. Comput. Theor. Nanosci. \textbf{10}, 1 (2013).
\bibitem{28}
 M. Cantoro, S. Hofmann, S. Pisana, V. Scardaci, A. Parvez, C. Ducati, A. C. Ferrari, A. M. Blackburn, K. Y. Wang, J. Robertson, Nano. Lett. \textbf{6}, 1107 (2006).
\bibitem{29}
R M. Sankaran, J. Phys. D: Appl. Phys.\textbf{44}, 174005 (2011).
\bibitem{30}
 F. Ding, P. Larsson, J. A. Larsson, R. Ahuja, H. Duan, A. Ros\'{e}n, K. Bolton, Nano Lett.\textbf{8}, 463 (2008).
\bibitem{31}
J.P. OByrne and Z. Li and  J. M. Tobin and J. A. Larsson and P. Larsson and R. Ahuja and J. D. Holmes
\bibitem{32}
 Q. Wang, S.-W. Yang, Y. Yang, M. B. Chan-Park, and Y. Chen, J. Phys. Chem. Lett. \textbf{2}, 1009 (2011).
\bibitem{33}
 We also optimized with the two triangles intact and a 5-atom ring was formed. 
\bibitem{34}
 L. Guan, K. Suenaga, and S. Iijima, Nano Lett. \textbf{2}, 459 (2008).
\bibitem{35}
 D. Takagi, Y. Kobayashi, and  Y. Homma,  J. Am. Chem. Soc. \textbf{131}, 6922 (2009).

 \end{thebibliography}
\end{document}